\documentclass[11pt]{article}

\usepackage[utf8]{inputenc}
\usepackage[T1]{fontenc}
\usepackage[margin=1in]{geometry}
\usepackage{amsmath,amssymb,amsthm,bm}
\usepackage{graphicx}
\usepackage{tikz}
\usetikzlibrary{shapes.geometric,arrows.meta,positioning,calc,backgrounds,fit}
\usepackage{enumerate}
\usepackage{natbib}
\usepackage{algorithm}
\usepackage{algpseudocode}
\usepackage[x11names]{xcolor}
\usepackage{multirow}
\usepackage{booktabs}
\usepackage{adjustbox}
\usepackage{microtype}
\usepackage[hidelinks]{hyperref}

\allowdisplaybreaks

\theoremstyle{plain}
\newtheorem{theorem}{Theorem}

\theoremstyle{definition}

\theoremstyle{remark}

\newcommand{\bd}{\boldsymbol}

\algnewcommand\INPUT{\item[\textbf{Input:}]}
\algnewcommand\OUTPUT{\item[\textbf{Output:}]}

\title{CRT*: Conditional Randomization Testing with Heterogeneous External and Unlabeled Data}
\author{Yingjie Zhang$^{1}$, Ziqi Chen$^{1,*}$, and Chenlei Leng$^{2,*}$\\[0.6em]
\small $^{1}$School of Statistics, KLATASDS-MOE, East China Normal University\\
\small $^{2}$Department of Applied Mathematics, Hong Kong Polytechnic University\\[0.4em]
\small $^{*}$Correspondence: zqchen@fem.ecnu.edu.cn; chenlei.leng@polyu.edu.hk}
\date{}

\begin{document}
\maketitle

\begin{abstract}
The conditional randomization test (CRT) provides a principled approach to conditional independence (CI) testing, guaranteeing exact type-I error control when the true conditional distribution is known. In practice, however, this distribution must be estimated, and estimation errors can inflate type-I errors, while high dimensionality and limited sample sizes can reduce power. Although external and unlabeled data offer the potential to improve CI testing, naive integration that ignores distributional heterogeneity can compromise type-I error control and fail to enhance power. We propose \textbf{CRT*}, a novel framework that robustly integrates external and unlabeled datasets to enhance CI testing in heterogeneous scenarios. CRT* employs smooth residual-bootstrap (SRB) with transfer learning for conditional distribution estimation, combined with adaptive data fusion via an optimal convex combination of test statistics. We theoretically establish that the SRB-based estimator converges to the true conditional distribution in expected total variation distance. Furthermore,  even in high-dimensional regimes, CRT* maintains valid type-I error control and achieves strictly higher power than standard CRT without external data. Simulations and RNA-seq breast cancer data analyses demonstrate that CRT* substantially improves power while maintaining type-I error control in  heterogeneous settings.
\end{abstract}


\section{Introduction}
\label{sec:intro}

Conditional independence (CI) testing is a foundational problem in statistics and machine learning, underpinning key advances in causal inference \citep{spirtes2000causation}, graphical models \citep{lauritzen1996graphical}, and variable selection \citep{dai2022significance}. We aim to determine whether two univariate random variables, $X$ and $Y$, are independent given a set of conditioning variables $Z \in \mathbb{R}^p$. Formally, the hypothesis is
\begin{equation*}
H_0: X \perp \!\!\! \perp Y|Z \quad \text{versus} \quad H_1: X \not \! \perp \!\!\! \perp Y|Z.
\end{equation*} 
Despite its fundamental importance, CI testing remains challenging due to limited sample sizes and the high dimensionality of the conditioning variables $Z$ \citep{shi2021double,li2023maxway,li2023nearest}. Among recent developments, the conditional randomization test (CRT) has emerged as a conceptually elegant approach that provides exact control of type-I error under idealized conditions \citep{candes2018panning}. However, significant practical hurdles remain, motivating further methodological development to ensure its validity and power in real-world settings.

\subsection{Idealized CRT}\label{subsec:utopia}

Under idealized conditions where the true conditional distribution of $X|Z$ is known, the CRT achieves exact type-I error control while permitting the use of any test statistic.
Specifically, suppose we observe $n$ i.i.d.\ samples $(X_i, Y_i, Z_i)_{i=1}^n$. Define $\bd X := (X_1, \ldots, X_n)^{\top}$, and similarly for $\bd Y$ and $\bd Z$. The CRT proceeds as follows:

\noindent
\textbf{1. Compute the test statistic on the observed dataset.} For a chosen test statistic $T$, compute $T(\bd X, \bd Y, \bd Z)$ on the observed data. 

\noindent
\textbf{2. Generate pseudo-samples.} For $m = 1, \ldots, M$, independently draw pseudo-samples $\bd X^{(m)} = (X_1^{(m)}, \ldots, X_n^{(m)})^{\top}$ from the conditional distribution $\rho_n(\cdot|\bd Z) = \prod_{i=1}^n \rho(\cdot|Z_i)$, where $\rho(\cdot|Z)$ denotes the true conditional distribution of $X|Z$.

\noindent
\textbf{3. Compute test statistics on the pseudo datasets.} For each pseudo dataset, compute the test statistic $T(\bd X^{(m)}, \bd Y, \bd Z)$. 

\noindent
\textbf{4. Calculate the $p$-value.} Let $\mathbb{I}\{\cdot\}$ be the indicator function. The $p$-value is defined as:
\begin{equation*}
p = \frac{1}{M+1} \left( 1 + \sum_{m=1}^M \mathbb{I}\left\{ T(\bd X^{(m)}, \bd Y, \bd Z) \geq T(\bd X, \bd Y, \bd Z) \right\} \right).
\end{equation*}

Under the null hypothesis $H_0$, the ($M+1$) triples $(\bd X, \bd Y, \bd Z), (\bd X^{(1)}, \bd Y, \bd Z)$, $\ldots$, $(\bd X^{(M)}, \bd Y, \bd Z)$ are exchangeable. Consequently, the $p$-value is valid in the sense that $\mathbb{P}_{H_0}(p \leq \alpha) \leq \alpha$ holds for any $\alpha \in [0,1]$ \citep{candes2018panning,berrett2020conditional}.

\subsection{CRT in Reality}\label{subsec:reality}

While the CRT is a powerful tool for controlling type-I error, its practical utility is often constrained by the requirement of knowing the true conditional distribution of $X|Z$. In fields such as genomics or the social sciences, this distribution is rarely known due to the high dimensionality of $Z$, the presence of non-Gaussian noise, and limited domain knowledge \citep{michal2024model,ismail2025deep}. 

In practice, researchers must estimate the conditional distribution $\rho_n(\cdot|\bd{Z})$ from data, yielding an estimator $\widehat{\rho}_n(\cdot|\bd{Z}) = \prod_{i=1}^n \widehat{\rho}(\cdot|Z_i)$ for resampling. This introduces estimation error, and the validity of the CRT then depends on the fidelity of $\widehat{\rho}_n$ to the truth. \citet{berrett2020conditional} demonstrated that such errors can inflate the type-I error, with the inflation bounded by the expected total variation (TV) distance between $\widehat{\rho}_n$ and $\rho_n$. However, \citet{shi2021double} showed that this expected TV distance may not vanish even as the sample size approaches infinity. 
To mitigate this, CRT variants leverage abundant auxiliary unlabeled data---where $X$ and $Z$ are observed but $Y$ is not---to improve estimation accuracy and achieve valid type-I error control \citep{berrett2020conditional,li2023maxway,yang2025conditional}. Nevertheless, heterogeneity between labeled and unlabeled datasets---arising from differences in populations or measurement protocols---coupled with the high dimensionality of $Z$, can substantially limit the effectiveness of these approaches.

Furthermore, the CRT often suffers from low power in settings with limited sample sizes and/or high-dimensional conditioning variables \citep{liu2022fast,li2024k,niu2024reconciling}. In summary, the practical implementation of the CRT faces two primary obstacles: (i) estimation error of the conditional distribution of $X|Z$, which can inflate the type-I error; and (ii) limited sample sizes and/or high dimensionality, which can reduce statistical power.

\subsection{CRT*}\label{subsec:crt*}

Recent advances in data science have significantly increased the availability of diverse datasets. In many applications, researchers have access not only to a primary internal labeled dataset but also to auxiliary unlabeled data and an external labeled cohort, the latter providing complete observations of $(X, Y, Z)$. These data sources offer new opportunities for CI testing: unlabeled samples can improve the estimation of the conditional distribution of $X|Z$, thereby facilitating valid type-I error control, while the external cohort provides additional information that can be leveraged to enhance testing power. However, effectively integrating heterogeneous data is non-trivial. We collectively refer to the internal and external datasets as the labeled datasets.

\smallskip
\noindent\textbf{Challenges in CI Testing with Multi-Source Data.}
We evaluated the performance of several state-of-the-art methods, including CRT (CRT1 and CRT2), CPT (CPT1 and CPT2) \citep{berrett2020conditional}, and Maxway CRT \citep{li2023maxway}, and compared them with our proposed \textbf{CRT*} under a nominal significance level $\alpha=0.05$. Further details are provided in Supplementary Materials~S.2. The results, summarized in Table~\ref{tab:ME}, highlight two critical findings: 
(i) \textbf{Heterogeneity between  internal and unlabeled datasets:} Existing methods ignore this heterogeneity when using the unlabeled data to estimate $\rho(\cdot|Z)$. Such estimators may be mismatched with the internal cohort, resulting in inflated type-I error. 
(ii) \textbf{Heterogeneity between  internal and external datasets:} If this heterogeneity is ignored and datasets are naively pooled for
CI testing using existing methods, the resulting tests may exhibit lower statistical power than when the external data were not used at all, rather than yielding the expected power gains. 

\begin{table}[htbp]
    \centering
    \small
    \begin{tabular}{lccc}
    \toprule
        \multirow{2}{*}{Method} & Type-I error & \multicolumn{2}{c}{Power}  \\
        \cmidrule(lr){3-4}
        & $n_{\text{E}}=0$ & $n_{\text{E}}=0$ & $n_{\text{E}}=200$\\
        \midrule
        CRT1 & 0.124 & 0.238 & 0.095 \\
        CRT2 & 0.067 & 0.240 & 0.124\\
        CPT1  & 0.229 & 0.036 & 0.039 \\
        CPT2 & 0.074 & 0.214 & 0.098 \\
        Maxway CRT & 0.163 & 0.266 & 0.146 \\
        \textbf{CRT*} & \textbf{0.044} & 0.230 & \textbf{0.687} \\
    \bottomrule
    \end{tabular}
    \caption{
    Type-I error under heterogeneity between internal and unlabeled datasets, and power under heterogeneity between internal and external datasets, for CRT, CPT, Maxway CRT, and CRT*.  $n_{\text{E}}$ denotes the external sample size.}
    \label{tab:ME}
\end{table}
The challenge of jointly leveraging multi-source data under heterogeneity raises a fundamental question: \emph{How can we robustly integrate heterogeneous unlabeled and external data to achieve valid and powerful CI testing in practice?} 

\smallskip
\noindent\textbf{Addressing Multi-Source Heterogeneity via CRT*.}
We propose \textbf{CRT*}, a novel framework for adaptive CI testing that robustly integrates unlabeled and external data  under distributional heterogeneity. Table \ref{tab:ME} shows that CRT* controls type-I error while effectively leveraging external data to substantially improve power. Our approach is motivated by 
the analysis of RNA-seq breast cancer data where the internal cohort (e.g., African American patients) is limited in size, while a larger external cohort (e.g., White patients) exhibits distributional heterogeneity. Furthermore, abundant unlabeled datasets (e.g., from the Gene Expression Omnibus) may exhibit conditional distributions different from those of both labeled cohorts (see Section~\ref{real data}).

To our knowledge, CRT* is the first framework designed to effectively unify internal, external, and unlabeled data for CI testing. Its key innovations include:
{\setlength{\leftmargini}{1em}
\begin{itemize}
    \item \emph{SRB with transfer learning for conditional distribution estimation:} CRT* leverages multi-source data via a novel smooth residual-bootstrap (SRB) framework integrated with transfer learning. This enables accurate estimation of the conditional distributions in both internal and external cohorts, ensuring robustness to heterogeneity between labeled and unlabeled sources. We further prove the consistency of our SRB-based conditional distribution estimators in expected total variation distance within high-dimensional regimes.
    \item \emph{Theoretical guarantees for type-I error:} We establish asymptotic type-I error control under the null, even when the external data only approximately satisfy conditional independence. 
    \item \emph{Adaptive data fusion:} CRT* constructs a fused test statistic using data-driven weights that explicitly account for distributional heterogeneity. This adaptive weighting prevents power loss and maximizes the benefit of the external cohort.
    \item \emph{Local asymptotic power analysis:} We derive power results that account for both distributional heterogeneity and estimation error. Importantly, we theoretically establish that CRT* achieves strictly higher power than standard CRT that does not incorporate external data.
    \item \emph{Empirical validation:} Extensive simulations and real-data analyses confirm that CRT* maintains valid type-I error control and increases power, even in high dimensions and under cross-cohort heterogeneity.
\end{itemize}}

\subsection{Related Work}
\label{related work}

A rich array of nonparametric methods has been developed for CI testing, encompassing metric-based, kernel-based, regression-based, and CRT-type approaches \citep{su2007consistent,fukumizu2007kernel,peters2014causal,candes2018panning,berrett2020conditional,liu2022fast,li2023maxway,zhang2024doubly}. Among these, the CRT and its variants are particularly attractive due to their rigorous type-I error control and methodological flexibility. To ensure validity, most CRT methods require that the conditional distribution of $X|Z$ be either known or accurately estimable from data that are distributionally matched to the target population. Furthermore, existing power analyses of the CRT are confined to narrow settings, typically relying on the assumption of a known conditional distribution or its first two moments \citep{wang2022high,katsevich2022power}. Crucially, current CRT methodologies do not leverage external data to improve power, nor do they address the challenges posed by distributional heterogeneity between internal and external datasets.

In contrast, the broader data fusion literature has primarily focused on parameter estimation, employing likelihood-based, convex-combination, or confidence-distribution methods to combine internal and external data \citep{lin2010relative,chatterjee2016constrained,shen2020fusion,gu2023meta}. These frameworks typically form statistics as convex combinations with weights chosen to minimize asymptotic variance, rather than to maximize power under the alternative in CI testing. How to effectively exploit external data to enhance the power of CI testing in heterogeneous scenarios, particularly in high-dimensional regimes, remains an open question.

Recent advances in transfer learning have demonstrated that auxiliary datasets can be leveraged to enhance the accuracy of parameter estimation \citep{li2022transfer,tian2023transfer,wang2023minimax,zhao2023residual}. Additionally, residual bootstrap procedures have been developed for conditional sampling without assuming a specific distributional form. However, existing methodology and theory are restricted to single-population settings \citep{freedman1981bootstrapping,neumeyer2009smooth}. To date, these disparate lines of work have not been unified into a framework that jointly leverages internal data alongside both unlabeled and external datasets for CI testing.

The remainder of the article is organized as follows. Section~\ref{transfer} introduces the CRT* framework. Section~\ref{simulation results} reports simulation studies, and Section~\ref{real data} demonstrates our methodology on RNA-seq breast cancer data. Section~\ref{conclusion} concludes with a discussion. Precise definitions of notation, all technical proofs, and  additional results are provided in the Supplementary Materials.

\section{CRT*: Data Fusion Conditional Randomization Test} \label{transfer}
We consider a setting where we observe $n$ internal samples $(X_i, Y_i, Z_i)_{i=1}^n$ and have access to $n_{\text{E}}$ external samples $(X_i^{\text{E}}, Y_i^{\text{E}}, Z_i^{\text{E}})_{i=1}^{n_{\text{E}}}$. Let $\bd X^{\text{E}} := (X_1^{\text{E}}, \ldots, X_{n_{\text{E}}}^{\text{E}})^{\top}$, with $\bd Y^{\text{E}}$ and $\bd Z^{\text{E}}$ defined analogously. The external data may originate from a population distinct from the internal one, potentially differing in both joint distribution and conditional dependence strength.  Throughout this work, we allow the common dimension of $Z$ and $Z^{\text{E}}$, denoted by $p$, to diverge with $n$. 

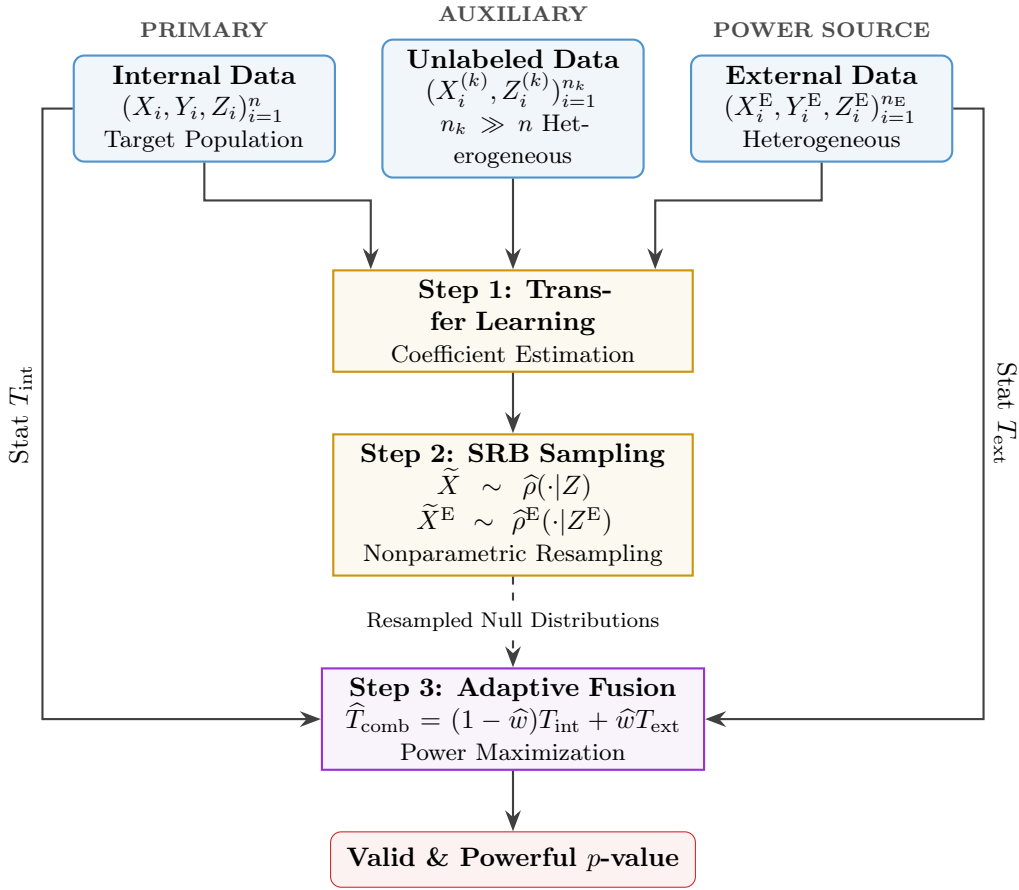
\begin{figure}[ht]
    \centering
    \begin{tikzpicture}[
        node distance=1.5cm and 0.5cm,
        data_box/.style={rectangle, draw=SteelBlue3, thick, rounded corners, fill=SteelBlue3!8, text width=3.2cm, align=center, minimum height=1.2cm, font=\small},
        proc_box/.style={rectangle, draw=DarkGoldenrod3, thick, fill=DarkGoldenrod3!6, text width=4.5cm, align=center, minimum height=1cm, font=\small},
        test_box/.style={rectangle, draw=DarkOrchid3, thick, fill=DarkOrchid3!6, text width=4.8cm, align=center, minimum height=1cm, font=\small},
        role_label/.style={font=\scriptsize\bfseries, color=gray!60!black},
        arrow/.style={-{Stealth[scale=1.2]}, thick, draw=gray!150},
        dashed_arrow/.style={-{Stealth[scale=1.2]}, thick, dashed, draw=gray!150}
    ]

        \node[data_box] (unlabeled) {
            \textbf{Unlabeled Data}\\
            $(X_i^{(k)}, Z_i^{(k)})_{i=1}^{n_k}$\\
            \footnotesize $n_k \gg n$ Heterogeneous
        };
        \node[role_label, above=0.1cm of unlabeled] {AUXILIARY};

        \node[data_box, left=0.6cm of unlabeled] (internal) {
            \textbf{Internal Data}\\
            $(X_i, Y_i, Z_i)_{i=1}^n$\\
            \footnotesize Target Population
        };
        \node[role_label, above=0.1cm of internal] {PRIMARY};

        \node[data_box, right=0.6cm of unlabeled] (external) {
            \textbf{External Data}\\
            $(X_i^{\text{E}}, Y_i^{\text{E}}, Z_i^{\text{E}})_{i=1}^{n_{\text{E}}}$\\
            \footnotesize Heterogeneous
        };
        \node[role_label, above=0.1cm of external] {POWER SOURCE};

        \node[proc_box, below=1.2cm of unlabeled] (trans) {
            \textbf{Step 1: Transfer Learning}\\
            \footnotesize Coefficient Estimation
        };

        \node[proc_box, below=0.8cm of trans] (srb) {
            \textbf{Step 2: SRB Sampling}\\
            $\widetilde{X} \sim \widehat{\rho}(\cdot| Z)$\\  $\widetilde{X}^{\text{E}} \sim \widehat{\rho}^{\text{E}}(\cdot| Z^{\text{E}})$\\
            \footnotesize Nonparametric Resampling
        };

        \node[test_box, below=1.2cm of srb] (fusion) {
            \textbf{Step 3: Adaptive Fusion}\\
            $\widehat{T}_{\text{comb}} = (1-\widehat{w})T_{\text{int}} + \widehat{w}T_{\text{ext}}$\\
            \footnotesize Power Maximization
        };

        \node[rectangle, rounded corners, draw=Firebrick3, fill=Firebrick3!6, below=0.8cm of fusion, font=\bfseries\small, inner sep=6pt] (output) {
            Valid \& Powerful $p$-value
        };

        \draw[arrow] (unlabeled) -- (trans);
        \draw[arrow] (internal.south) -- ++(0,-0.5) -| (trans.160);
        \draw[arrow] (external.south) -- ++(0,-0.5) -| (trans.20);
        
        \draw[arrow] (trans) -- (srb);

        \draw[dashed_arrow] (srb) -- (fusion) 
            node[midway, fill=white, font=\scriptsize, inner sep=2pt] {Resampled Null Distributions};

        \draw[arrow] (internal.west) -- ++(-0.4,0) |- (fusion.west)
            node[pos=0.25, left, font=\small, rotate=90, anchor=south] {Stat $T_{\text{int}}$};
        \draw[arrow] (external.east) -- ++(0.4,0) |- (fusion.east)
            node[pos=0.25, right, font=\small, rotate=-90, anchor=south] {Stat $T_{\text{ext}}$};

        \draw[arrow] (fusion) -- (output);

    \end{tikzpicture}
    \caption{The \textbf{CRT*} Workflow: Leveraging unlabeled data for robust distribution estimation via SRB with transfer learning, while utilizing external data for adaptive power enhancement through optimal weighting.}
    \label{fig:0}
\end{figure}

This section develops the new \textbf{CRT*} framework. Specifically,  we propose the smooth residual-bootstrap approach with transfer learning to robustly estimate the conditional distributions and enable valid CI testing (see Sections \ref{A transfer learning with smooth residual bootstrap based CRT*} and \ref{CRT1 Controls Type-I Error}). Furthermore, to enhance testing power, we use adaptive data fusion through convex combinations of statistics to incorporate external dataset (see Section \ref{CRT_power}). A conceptual framework and workflow for our proposed  CRT*  is provided in Figure~\ref{fig:0}.

\subsection{CRT* with Estimated Conditional Distributions}\label{A transfer learning with smooth residual bootstrap based CRT*}

As previously discussed, accurately estimating the conditional distribution of $\bd X|\bd Z$, where $\bd X=(X_1,\ldots,X_n)^{\top}$ and $\bd Z=(Z_1,\ldots,Z_n)^{\top}$, is infeasible using only the internal data, and this task becomes increasingly difficult in high-dimensional regimes with limited sample sizes. We assume access to $K$ auxiliary unlabeled datasets, where the $k$-th dataset contains $n_k$ samples $(X_i^{(k)}, Z_i^{(k)})_{i=1}^{n_k}$ for $k=1,\ldots,K$, with $K$ fixed. Let $\bd{X}^{(k)} := (X^{(k)}_1,\ldots,X^{(k)}_{n_k})^{\top}$ and $\bd{Z}^{(k)} := (Z^{(k)}_1,\ldots,Z^{(k)}_{n_k})^{\top}$. Since unlabeled datasets are often abundant in practice \citep{yang2022survey}, we leverage these auxiliary data together with the labeled datasets to learn the conditional distributions required for resampling in CRT*.

However, the conditional distribution of $X|Z$ often differs between labeled and unlabeled datasets due to variations in data collection protocols, population characteristics, or experimental conditions. For instance, in our gene expression analysis, labeled data are drawn from the TCGA-BRCA cohort via the UCSC Xena platform, while unlabeled data are sourced from the Gene Expression Omnibus (GEO) database. Such heterogeneity makes it challenging to directly apply models trained on unlabeled data to the labeled cohort. To address this, transfer learning methods are indispensable, as they are specifically designed to adapt to discrepancies between the labeled and unlabeled datasets \citep{li2022transfer,zhao2023residual}.

In gene expression studies, high-dimensional sparse linear regression is a natural and effective framework for modeling the relationship between a target log-transformed gene expression $X$ and other log-transformed gene expressions $Z$ \citep{fan2009network,choi2017poisson,li2022transfer}. Accordingly, we assume the internal, external, and auxiliary unlabeled datasets follow the high-dimensional linear models:
\begin{align}
\bd{X} &= \bd{Z}\bd{\beta} + \bd{\epsilon}, \quad \bd{Z} \perp \!\!\! \perp \bd{\epsilon}, \label{primary internal data model}  \\   
\bd{X}^{\text{E}} &= \bd{Z}^{\text{E}}\bd{\beta}^{\text{E}} + \bd{\epsilon}^{\text{E}}, \quad \bd{Z}^{\text{E}} \perp \!\!\! \perp \bd{\epsilon}^{\text{E}}, \label{primary external data model} \\
\bd{X}^{(k)} &= \bd{Z}^{(k)}\bd{w}^{(k)} + \bd{\epsilon}^{(k)}, \quad \bd{Z}^{(k)} \perp \!\!\! \perp \bd{\epsilon}^{(k)}, \quad k=1,\ldots,K, \label{unlabel data model}
\end{align}
where $\bd{\epsilon}$, $\bd{\epsilon}^{\text{E}}$, and $\bd{\epsilon}^{(k)}$ are vectors of i.i.d. noise terms. We assume the regression coefficients $\bd{\beta}$, $\bd{\beta}^{\text{E}}$, and $\bd{w}^{(k)}$ are $p$-dimensional sparse vectors that may vary across datasets, satisfying $\|\bd{\beta}\|_0 \leq s$, $\|\bd{\beta}^{\text{E}}\|_0 \leq s^{\text{E}}$, and $\|\bd{w}^{(k)}\|_0 \leq s^{(k)}$, where $s, s^{\text{E}}, s^{(k)} \ll p$. Crucially, all noise terms are assumed to be sub-Gaussian with mean zero and variance $\widetilde{\sigma}^2$, which relaxes the Gaussianity assumptions prevalent in prior CRT literature \citep{wang2022high,li2023maxway}.

CRT* requires precise sampling from the conditional distribution of $X|Z$. When the noise distribution is unknown or non-Gaussian, this task becomes substantially more challenging. To overcome this, we combine the smooth residual-bootstrap (SRB) method \citep{neumeyer2009smooth} with transfer learning. This integrated approach leverages heterogeneous unlabeled data to produce accurate, nonparametric approximations of the conditional distributions, enabling valid resampling without imposing restrictive parametric forms on the noise.

Specifically, we construct ``informative sets'' of auxiliary samples whose regression coefficients are similar to those of the internal data:
\begin{equation}\label{internal informative set}
    \mathcal{A} = \left\{ 1 \leq k \leq K : \|\bd{\beta} - \bd{w}^{(k)}\|_1 \leq h \right\}, \qquad N_{\mathcal{A}} = \sum_{k \in \mathcal{A}} n_k.
\end{equation}
Similarly, for the external data, we define:
\begin{equation}\label{external informative set}
    \mathcal{A}^{\text{E}} = \left\{ 1 \leq k \leq K : \|\bd{\beta}^{\text{E}} - \bd{w}^{(k)}\|_1 \leq h^{\text{E}} \right\}, \qquad N_{\mathcal{A}^{\text{E}}} = \sum_{k \in \mathcal{A}^{\text{E}}} n_k.
\end{equation}
Here, $h$ and $h^{\text{E}}$ characterize the  discrepancy between the regression coefficients of the  unlabeled samples and the labeled datasets. Smaller values indicate higher similarity. 

Our procedure first employs the oracle Trans-Lasso \citep{li2022transfer}, detailed in Algorithm~S.1 in Supplementary Materials~S.12, to integrate information from both the informative auxiliary unlabeled samples and the labeled datasets, producing debiased estimates of the regression coefficients. We then apply the SRB for conditional sampling based on the estimated residuals. The full procedure is summarized in four steps below (see Figure~\ref{fig: flowchart} for a flowchart and Algorithm~S.2 in Supplementary Materials~S.12 for pseudocode).

\noindent\textbf{Step 1: Data splitting.}  
Partition the  labeled datasets into disjoint subsets. Let $\mathcal{I}_{\text{I}} \subset \{1,\ldots,n\}$ with $|\mathcal{I}_{\text{I}}| = \lfloor w_1 n \rfloor$ and $\mathcal{I}_{\text{E}} \subset \{1,\ldots,n_{\text{E}}\}$ with $|\mathcal{I}_{\text{E}}| = \lfloor w_2 n_{\text{E}} \rfloor$, where $w_1, w_2 \in (0,1)$. The hold-out subsets $\mathcal{I}_{\text{I}}$ and $\mathcal{I}_{\text{E}}$ are reserved for computing test statistics, while the complements $\mathcal{I}_{\text{I}}^c$ and $\mathcal{I}_{\text{E}}^c$, along with the informative auxiliary unlabeled datasets, are used for conditional-distribution estimation.

\noindent\textbf{Step 2: Coefficient estimation via transfer learning.}  
Apply oracle Trans-Lasso (Algorithm~S.1) to estimate $\bd{\beta}$ using $(\bd{X}_{\mathcal{I}_{\text{I}}^c}, \bd{Z}_{\mathcal{I}_{\text{I}}^c})$ as the target data and $\{(\bd{X}^{(k)}, \bd{Z}^{(k)})\}_{k \in \mathcal{A}}$ as the source data. Similarly, estimate $\bd{\beta}^{\text{E}}$ using $(\bd{X}_{\mathcal{I}_{\text{E}}^c}^{\text{E}}, \bd{Z}_{\mathcal{I}_{\text{E}}^c}^{\text{E}})$ and $\{(\bd{X}^{(k)}, \bd{Z}^{(k)})\}_{k \in \mathcal{A}^{\text{E}}}$. Let the resulting estimators be $\widehat{\bd{\beta}}$ and $\widehat{\bd{\beta}}^{\text{E}}$. For each $k \in \mathcal{A} $,  estimate $\bd{w}^{(k)}$ using oracle Trans-Lasso, treating $(\bd{X}^{(k)}, \bd{Z}^{(k)})$ as target and $(\bd{X}_{\mathcal{I}_{\text{I}}^c},\bd{Z}_{\mathcal{I}_{\text{I}}^c})\cup\{(\bd{X}^{(j)},\bd{Z}^{(j)})\}_{j\in \mathcal{A}\setminus \{k\}}$ as source. 
Similarly, for each $k \in \mathcal{A}^{\text{E}}$, estimate $\bd{w}^{(k)}$ using $(\bd{X}^{(k)},\bd{Z}^{(k)})$ and $(\bd{X}_{\mathcal{I}_{\text{E}}^c}^{\text{E}},\bd{Z}_{\mathcal{I}_{\text{E}}^c}^{\text{E}})\cup\{(\bd{X}^{(j)},\bd{Z}^{(j)})\}_{j\in \mathcal{A}^{\text{E}}\setminus \{k\}}$. Denote the resulting estimators by $\{\widehat{\bd w}^{(k)}\}_{k\in \mathcal{A}}$ and $\{\widehat{\bd w}^{(k)}_{\text{E}}\}_{k\in \mathcal{A}^{\text{E}}}$. 


\noindent\textbf{Step 3: Smooth residual-bootstrap (SRB) conditional sampling.}  
Compute the residuals for the labeled and informative auxiliary unlabeled datasets:
\begin{equation}\label{hatepsilon} 
    \widehat{\bd{\epsilon}}_{\mathcal{I}_{\text{I}}^c} = \bd{X}_{\mathcal{I}_{\text{I}}^c} - \bd{Z}_{\mathcal{I}_{\text{I}}^c} \widehat{\bd{\beta}}, \quad
    \widehat{\bd{\epsilon}}^{\text{E}}_{\mathcal{I}_{\text{E}}^c} = \bd{X}^{\text{E}}_{\mathcal{I}_{\text{E}}^c} - \bd{Z}^{\text{E}}_{\mathcal{I}_{\text{E}}^c} \widehat{\bd{\beta}}^{\text{E}}, 
\end{equation}
\begin{equation}\label{unlabelhatepsilon}
    \widehat{\bd{\epsilon}}^{(k)} = \bd{X}^{(k)} - \bd{Z}^{(k)} \widehat{\bd{w}}^{(k)}~(k \in \mathcal{A}), \quad
    \widehat{\bd{\epsilon}}^{\text{E}^{(k)}} = \bd{X}^{(k)} - \bd{Z}^{(k)} \widehat{\bd{w}}^{(k)}_{\text{E}}~(k \in \mathcal{A}^{\text{E}}).
\end{equation}
Concatenate the estimated residuals:
$$
    \widehat{\bd{\epsilon}}^{\text{A}} = \left[\widehat{\bd{\epsilon}}_{\mathcal{I}_{\text{I}}^c}^{\top}, (\widehat{\bd{\epsilon}}^{(k_1)})^{\top}, \ldots, (\widehat{\bd{\epsilon}}^{(k_{|\mathcal{A}|})})^{\top}\right]^{\top}, ~~
    \widehat{\bd{\epsilon}}^{\text{E}^{\text{A}}}=\Big[(\widehat{\bd{\epsilon}}_{\mathcal{I}_{\text{E}}^c}^{\text{E}})^{\top}, (\widehat{\bd{\epsilon}}^{\text{E}^{(k_1^{\text{E}})}})^{\top}, \ldots, (\widehat{\bd{\epsilon}}^{\text{E}^{(k_{|\mathcal{A}^{\text{E}}|}^{\text{E}})}})^{\top}\Big]^{\top},
$$
where $\left\{k_1,\ldots,k_{|\mathcal{A}|}\right\}=\mathcal{A}$ and $\left\{k_1^{\text{E}},\ldots,k_{|\mathcal{A}^{\text{E}}|}^{\text{E}}\right\}=\mathcal{A}^{\text{E}}$. Center the residuals:
\begin{align}
    \epsilon_i' &= \widehat{\epsilon}_i^{\text{A}} - \frac{1}{|\mathcal{I}_{\text{I}}^c| + N_{\mathcal{A}}} \left( \sum_{j \in \mathcal{I}_{\text{I}}^c} \widehat{\epsilon}_j + \sum_{k \in \mathcal{A}} \sum_{j=1}^{n_k} \widehat{\epsilon}_j^{(k)} \right), \label{center internal residual} \\
    \epsilon_i^{\text{E}'} &= \widehat{\epsilon}_i^{\text{E}^{\text{A}}} - \frac{1}{|\mathcal{I}_{\text{E}}^c| + N_{\mathcal{A}^{\text{E}}}} \left( \sum_{j \in \mathcal{I}_{\text{E}}^c} \widehat{\epsilon}_j^{\text{E}} + \sum_{k \in \mathcal{A}^{\text{E}}} \sum_{j=1}^{n_k} \widehat{\epsilon}_j^{\text{E}^{(k)}} \right). \label{center external residual}
\end{align}
For $m = 1, \ldots, M$, draw bootstrap residuals $\{\epsilon_i''^{(m)}\}_{i \in \mathcal{I}_{\text{I}}}$ and $\{\epsilon_i^{\text{E}''^{(m)}}\}_{i \in \mathcal{I}_{\text{E}}}$ with replacement from $\{\epsilon_i'\}_{i=1}^{|\mathcal{I}_{\text{I}}^c| + N_{\mathcal{A}}}$ and $\{\epsilon_i^{\text{E}'}\}_{i=1}^{|\mathcal{I}_{\text{E}}^c| + N_{\mathcal{A}^{\text{E}}}}$, respectively. Generate smooth bootstrap residuals $\widetilde{\epsilon}_i^{(m)} = \epsilon_i''^{(m)} + a_n U_i^{(m)}$ and $\widetilde{\epsilon}_i^{\text{E}^{(m)}} = \epsilon_i^{\text{E}''^{(m)}} + a_{n_{\text{E}}} U_i^{\text{E}^{(m)}}$, where $a_n, a_{n_{\text{E}}} > 0$ are smoothing parameters, and $U_i^{(m)}, U_i^{\text{E}^{(m)}}$ are i.i.d. standard normal random variables. Construct pseudo-samples:
\begin{align*}
    \widetilde{\bd{X}}_{\mathcal{I}_{\text{I}}}^{(m)} = \bd{Z}_{\mathcal{I}_{\text{I}}} \widehat{\bd{\beta}} + \widetilde{\bd{\epsilon}}_{\mathcal{I}_{\text{I}}}^{(m)}, \quad
    \widetilde{\bd{X}}_{\mathcal{I}_{\text{E}}}^{\text{E}^{(m)}} = \bd{Z}^{\text{E}}_{\mathcal{I}_{\text{E}}} \widehat{\bd{\beta}}^{\text{E}} + \widetilde{\bd{\epsilon}}^{\text{E}^{(m)}}_{\mathcal{I}_{\text{E}}}.
\end{align*}

\noindent\textbf{Step 4: $p$-value computation.}  
For a given test statistic $T$, compute the $p$-value:
\begin{align*}
    p = \frac{1}{M+1} \left( 1 + \sum_{m=1}^M \mathbb{I} \left\{ T(\widetilde{\bd{X}}_{\mathcal{I}_{\text{C}}}^{\text{C}^{(m)}}, \bd{Y}_{\mathcal{I}_{\text{C}}}^{\text{C}}, \bd{Z}_{\mathcal{I}_{\text{C}}}^{\text{C}}) \geq T(\bd{X}_{\mathcal{I}_{\text{C}}}^{\text{C}}, \bd{Y}_{\mathcal{I}_{\text{C}}}^{\text{C}}, \bd{Z}_{\mathcal{I}_{\text{C}}}^{\text{C}}) \right\} \right),
\end{align*}
where $\bd{X}_{\mathcal{I}_{\text{C}}}^{\text{C}} = \left[ (\bd{X}_{\mathcal{I}_{\text{I}}})^{\top}, (\bd{X}_{\mathcal{I}_{\text{E}}}^{\text{E}})^{\top} \right]^{\top}$, with $\bd{Y}_{\mathcal{I}_{\text{C}}}^{\text{C}}, \bd{Z}_{\mathcal{I}_{\text{C}}}^{\text{C}}$, and $\widetilde{\bd{X}}_{\mathcal{I}_{\text{C}}}^{\text{C}^{(m)}}$ defined analogously.

By leveraging auxiliary unlabeled datasets via transfer learning, we obtain accurate coefficient estimates and well-estimated residuals even in high-dimensional regimes with limited labeled data. The smoothing step in the SRB is essential: it ensures the estimated conditional distributions are  continuous, allowing the total variation (TV) distance to the truth to vanish asymptotically. This prevents the type-I error inflation typically seen with discrete resampling (see Section~\ref{CRT1 Controls Type-I Error} and Supplementary Materials~S.13.6).

\begin{figure}[!t]
\centering
\includegraphics[width=\linewidth,trim=90 0 230 0,clip]{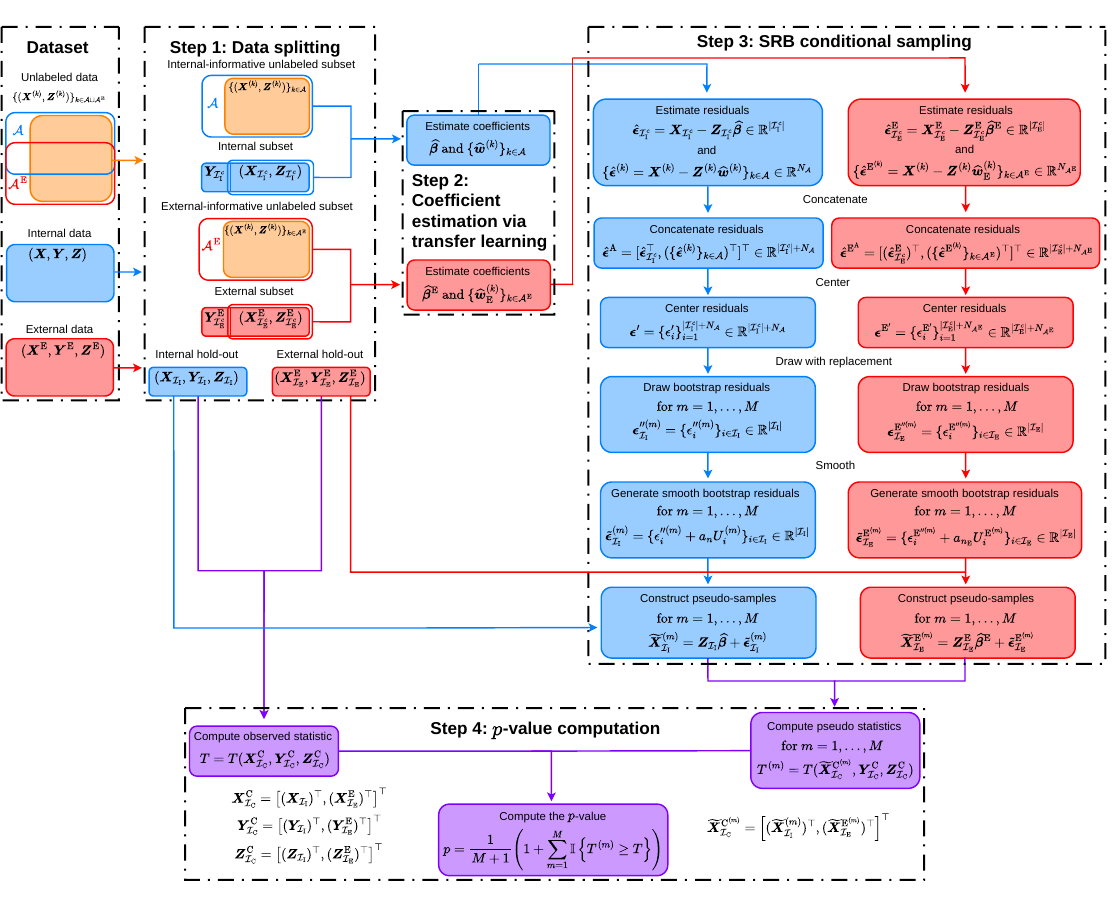}
\caption{Flowchart of the CRT* procedure with a given test statistic $T$. The power-enhancing test statistic  is defined in (\ref{IRW stat}).}
\label{fig: flowchart}
\end{figure}

\subsection{CRT* Controls Type-I Error}\label{CRT1 Controls Type-I Error}

We next establish the asymptotic type-I error control for the CRT* procedure. In our framework, we use the datasets $(\bd{X}_{\mathcal{I}_{\text{I}}^c}, \bd{Z}_{\mathcal{I}_{\text{I}}^c})$ and $\{(\bd{X}^{(k)}, \bd{Z}^{(k)})\}_{k \in \mathcal{A}}$ via SRB with transfer learning to construct an estimator $\widehat{\rho}_{|\mathcal{I}_{\mathrm{I}}|}(\cdot|\bd{Z}_{\mathcal{I}_{\mathrm{I}}}) = \prod_{i\in \mathcal{I}_{\mathrm{I}}}\widehat{\rho}(\cdot|Z_i)$ of the true conditional distribution $\rho_{|\mathcal{I}_{\mathrm{I}}|}(\cdot|\bd{Z}_{\mathcal{I}_{\mathrm{I}}})$. Similarly, we construct an estimator $\widehat{\rho}^{\text{E}}_{|\mathcal{I}_{\mathrm{E}}|}(\cdot|\bd{Z}^{\text{E}}_{\mathcal{I}_{\mathrm{E}}})$ of the true conditional distribution ${\rho}^{\text{E}}_{|\mathcal{I}_{\mathrm{E}}|}(\cdot|\bd{Z}^{\text{E}}_{\mathcal{I}_{\mathrm{E}}})$ for the external data. 
Let 
$\rho_{|\mathcal{I}_{\mathrm{E}}|}^{\mathrm{E}}(\cdot|\bd{Y}_{\mathcal{I}_{\mathrm{E}}}^{\mathrm{E}}, \bd{Z}_{\mathcal{I}_{\mathrm{E}}}^{\mathrm{E}})$ 
denote the conditional distribution of $\bd{X}_{\mathcal{I}_{\mathrm{E}}}^{\mathrm{E}}$ given $(\bd{Y}_{\mathcal{I}_{\mathrm{E}}}^{\mathrm{E}}, \bd{Z}_{\mathcal{I}_{\mathrm{E}}}^{\mathrm{E}})$.

\begin{theorem}\label{asymptotic valid SSL-CRT1}
For any $\alpha \in [0,1]$, the $p$-value of $\mathrm{CRT}^*$ satisfies:
\begin{equation}\label{first claim of T1}
    \mathbb{P}_{H_0}\left(p \leq \alpha \mid \bd Y_{\mathcal{I}_{\mathrm{C}}}^{\mathrm{C}}, \bd Z_{\mathcal{I}_{\mathrm{C}}}^{\mathrm{C}} \right)
    \leq \alpha + d_{\mathrm{TV}}\left(
        \rho_{|\mathcal{I}_{\mathrm{I}}|} \cdot \rho_{|\mathcal{I}_{\mathrm{E}}|}^{\mathrm{E}}(\cdot|\bd{Y}_{\mathcal{I}_{\mathrm{E}}}^{\mathrm{E}}, \bd{Z}_{\mathcal{I}_{\mathrm{E}}}^{\mathrm{E}})~,~
        \widehat{\rho}_{|\mathcal{I}_{\mathrm{I}}|} \cdot \widehat{\rho}^{\mathrm{E}}_{|\mathcal{I}_{\mathrm{E}}|}
    \right).
\end{equation}
Marginalizing over $\bd Y_{\mathcal{I}_{\mathrm{C}}}^{\mathrm{C}}$ and $\bd Z_{\mathcal{I}_{\mathrm{C}}}^{\mathrm{C}}$ yields:
\begin{equation}\label{tv bound of OTLASRBout}
\begin{aligned}
    \mathbb{P}_{H_0}\left(p\leq \alpha \right)
    \leq \alpha
    &+ \sqrt{\frac{1}{2} |\mathcal{I}_{\mathrm{E}}| \cdot I(X^{\mathrm{E}}; Y^{\mathrm{E}}|Z^{\mathrm{E}})} \\
    &+ \mathbb{E}\left[ d_{\mathrm{TV}}\left(\rho_{|\mathcal{I}_{\mathrm{I}}|}, \widehat{\rho}_{|\mathcal{I}_{\mathrm{I}}|}\right) \right] + \mathbb{E}\left[ d_{\mathrm{TV}}\left(\rho^{\mathrm{E}}_{|\mathcal{I}_{\mathrm{E}}|}(\cdot|\bd{Z}^{\mathrm{E}}_{\mathcal{I}_{\mathrm{E}}}), \widehat{\rho}^{\mathrm{E}}_{|\mathcal{I}_{\mathrm{E}}|}\right) \right], 
\end{aligned}
\end{equation}
where $I(X^{\mathrm{E}}; Y^{\mathrm{E}}|Z^{\mathrm{E}})$ is the conditional mutual information of the external data.
\end{theorem}

Theorem~\ref{asymptotic valid SSL-CRT1} holds for any test statistic (the proof is provided in Supplementary Materials S.11). In \eqref{tv bound of OTLASRBout}, the second term quantifies the deviation from the null hypothesis in the external data, while the third and fourth terms measure the estimation accuracy of the conditional distributions for the internal and external cohorts, respectively. If $I(X^{\mathrm{E}}; Y^{\mathrm{E}}|Z^{\mathrm{E}}) = o(1/|\mathcal{I}_{\mathrm{E}}|)$  (see Supplementary Materials~S.6.1 for interpretation) and the expected total variation (TV) distances vanish as sample sizes increase, CRT* asymptotically controls the type-I error at level $\alpha$. Notably, our framework permits the external dataset to exhibit weak conditional dependence while maintaining valid error control under the internal null. Moreover, it accommodates  distributional heterogeneity, allowing the joint distributions of $(X,Y,Z)$ and $(X^{\text{E}}, Y^{\text{E}}, Z^{\text{E}})$ to differ.
We further provide a lower bound on type-I error inflation in Supplementary Materials~S.3, demonstrating that
the bound in (\ref{first claim of T1}) is asymptotically tight.


This result generalizes Theorem 4 of \cite{berrett2020conditional} by incorporating external data. 
When no external data are available, the second and fourth terms vanish, and the bound recovers that in Theorem 4 of \cite{berrett2020conditional}.

To establish that the TV terms in Theorem~\ref{asymptotic valid SSL-CRT1} converge to zero, we impose the following standard conditions:
{\setlength{\leftmargini}{2em}
\begin{enumerate}
\renewcommand{\labelenumi}{(A\arabic{enumi})}
\renewcommand{\theenumi}{A\arabic{enumi}}

    \item \label{Gaussian design} ~$Z$, $Z^{\text{E}}$, and $Z^{(k)}$ (for $k \in \mathcal{A} \cup \mathcal{A}^{\text{E}}$) are i.i.d.\ Gaussian with mean zero and covariance $\bd{\Sigma}$, where the eigenvalues of $\bd{\Sigma}$ are bounded away from zero and infinity.
\item \label{second moment bound} ~The second moments of $X$, $X^{\text{E}}$, and $X^{(k)}$ (for $k \in \mathcal{A} \cup \mathcal{A}^{\text{E}}$) are finite. The noise terms are i.i.d.\ sub-Gaussian with mean zero and variance $0 < \widetilde{\sigma}^2 < \infty$, and their common density satisfies certain regularity conditions (see Supplementary Materials~S.4).

\end{enumerate}
}

Under the conditions stated above, with regularization parameters as specified in Theorem S.2 in Supplementary Materials S.4, and assuming that the smoothing parameter satisfies $a_n \to 0$, we establish the convergence rate for the internal conditional distribution estimator. Specifically, we have:
\begin{equation*}
    \mathbb{E}\left[d_{\mathrm{TV}}\left(\widehat{\rho}_{|\mathcal{I}_{\mathrm{I}}|}, \rho_{|\mathcal{I}_{\mathrm{I}}|}\right) \mid \mathcal{D}_{\mathcal{A}}\right]
    = O_{\mathbb{P}}\left(a_n\sqrt{|\mathcal{I}_{\mathrm{I}}|} + \frac{\sqrt{|\mathcal{I}_{\mathrm{I}}|}}{a_n} \cdot \sqrt{\frac{\widetilde{s}\log p}{|\mathcal{I}_{\mathrm{I}}^c|+N_{\mathcal{A}}} + \frac{\widetilde{s}\log p}{\widetilde{n}} \wedge \widetilde{\eta}_h} \right),
\end{equation*}
where $\mathcal{D}_{\mathcal{A}} = (\bd{X}_{\mathcal{I}_{\text{I}}^c}, \bd{Z}_{\mathcal{I}_{\text{I}}^c}) \cup \{(\bd{X}^{(k)}, \bd{Z}^{(k)})\}_{k\in \mathcal{A}}$ represents the data used to estimate the internal conditional distribution. Here, $\widetilde{s} = s \vee \max\{s^{(k)} : k \in \mathcal{A}\}$ denotes the effective sparsity, $\widetilde{n} = |\mathcal{I}_{\text{I}}^c| \wedge \min\{n_k : k \in \mathcal{A}\}$ is the minimum cohort size, and $\widetilde{\eta}_{h} = \left(h\sqrt{\log p / \widetilde{n}}\right) \wedge h^2$ accounts for the heterogeneity discrepancy. 

The factor $\sqrt{|\mathcal{I}_{\text{I}}|}$ in the convergence rate arises because both $\widehat{\rho}_{|\mathcal{I}_{\mathrm{I}}|}$ and $\rho_{|\mathcal{I}_{\mathrm{I}}|}$ are products of $|\mathcal{I}_{\mathrm{I}}|$ conditional distributions. Moreover, the  rate admits a clear bias-variance decomposition: the term $a_n\sqrt{|\mathcal{I}_{\mathrm{I}}|}$ represents the bias introduced by smoothing, while the second term captures the variance contribution from estimating the high-dimensional regression coefficients and the residual distribution. By balancing these terms with the choice $a_n\asymp \left( \frac{\widetilde{s}\log p}{|\mathcal{I}_{\mathrm{I}}^c| + N_{\mathcal{A}}} + \frac{\widetilde{s}\log p}{\widetilde{n}} \wedge \widetilde{\eta}_h \right)^{1/4},$
we obtain the optimal convergence rate of $\sqrt{|\mathcal{I}_{\mathrm{I}}|}\left( \frac{\widetilde{s}\log p}{|\mathcal{I}_{\mathrm{I}}^c| + N_{\mathcal{A}}} + \frac{\widetilde{s}\log p}{\widetilde{n}} \wedge \widetilde{\eta}_h \right)^{1/4}$. This rate vanishes asymptotically if $h = o(|\mathcal{I}_{\text{I}}|^{-1})$ and the total unlabeled sample size $N_{\mathcal{A}}$ is sufficiently large (specifically $N_{\mathcal{A}} \gg |\mathcal{I}_{\text{I}}|^2 \widetilde{s}\log p$). Such requirements are easily satisfied in practice, as auxiliary unlabeled data are typically abundant. These results hold analogously for the external data. Therefore, Theorem~S.2 establishes that the CRT*  asymptotically controls the type-I error:
\begin{equation*}
    \mathbb{P}_{H_0}\left(p \leq \alpha \right) \leq \alpha + o(1), \quad \forall \alpha \in [0,1]. 
\end{equation*}


To our knowledge, this is the first result establishing convergence in TV distance for a residual-bootstrap-based conditional distribution estimator. While existing work has demonstrated convergence in the Mallows metric \citep{freedman1981bootstrapping,lopes2014residual}, Mallows convergence does not imply TV convergence. Crucially, the unsmoothed residual-bootstrap distribution is discrete, whereas the true distribution is typically continuous; consequently, their TV distance remains equal to one for any sample size. By incorporating the smoothing step, our SRB framework ensures that the estimated distribution is  continuous, allowing the TV distance to vanish asymptotically. This property is fundamental to the validity of the CRT* framework.

\subsection{CRT* Increases Power}\label{CRT_power}

While Section~\ref{CRT1 Controls Type-I Error} establishes that CRT* asymptotically controls type-I error for any test statistic, the choice of statistic is critical for achieving high statistical power. Naively pooling internal and external data---without accounting for distributional differences---not only fails to improve power but can even substantially diminish it, as demonstrated by the motivating results in Table~\ref{tab:ME} and the detailed simulations in Sections~\ref{KIS,DM,SC} and \ref{KIS,DM,DC}.

To address this and enhance power, we utilize a convex combination of \textit{distilled statistics} \citep{liu2022fast}, which are specifically designed to handle heterogeneity across cohorts. A distilled statistic is constructed by first estimating the conditional expectations $g(Z) = \mathbb{E}(Y|Z)$ and $\mu(Z) = \mathbb{E}(X|Z)$ with estimators $\widehat{g}$ and $\widehat{\mu}$, respectively, and then computing the sample covariance between the corresponding residuals. For the internal dataset, we define:
\begin{equation}
    T_{\text{int}} = \frac{1}{|\mathcal{I}_{\mathrm{I}}|}\sum_{i\in \mathcal{I}_{\mathrm{I}}} \left(Y_i - \widehat{g}(Z_i)\right)\left(X_i - \widehat{\mu}(Z_i)\right),\label{distilled internal}
\end{equation}
and analogously for the external dataset:
\begin{equation}
    T_{\text{ext}} = \frac{1}{|\mathcal{I}_{\mathrm{E}}|}\sum_{i\in \mathcal{I}_{\mathrm{E}}} \left(Y_i^{\text{E}} - \widehat{g}_{\text{E}}(Z_i^{\text{E}})\right)\left(X_i^{\text{E}} - \widehat{\mu}_{\text{E}}(Z_i^{\text{E}})\right),\label{distilled external}
\end{equation}
where $\widehat{g}_{\text{E}}$ and $\widehat{\mu}_{\text{E}}$ are estimators of $\mathbb{E}(Y^{\text{E}}|Z^{\text{E}})$ and $\mathbb{E}(X^{\text{E}}|Z^{\text{E}})$, respectively. To combine information from both datasets while accommodating distributional heterogeneity, we propose the combined statistic:
\begin{equation}\label{eq:combined_ds2}
    T_{\text{comb}} = (1-w) T_{\text{int}} + w T_{\text{ext}},
\end{equation}
where $w \in [0,1]$ is a weight parameter that can be adaptively tuned to maximize testing power.

In this work, we focus on the two-sided test statistic $|T_{\text{comb}}|$, while the power analysis for the one-sided version is provided in Supplementary Materials~S.10. Unlike naive pooling, which computes a single distilled statistic on a concatenated dataset, our approach explicitly respects cohort-specific distributions by constructing separate statistics before fusion. This methodology ensures that heterogeneity across cohorts is leveraged as a source of information rather than a source of bias, thereby maximizing the potential for power gains from external data.

For the theoretical analysis of testing power, we focus on local alternatives \citep{wang2022high,katsevich2022power, niu2024reconciling}. We introduce the high-dimensional partial linear models:
\begin{align}
    &Y = cX + \bar{g}(Z) + \epsilon_y, \quad (X, Z) \perp\!\!\!\perp \epsilon_y, \label{eq:model_internal} \\
    &Y^{\text{E}} = c^{\text{E}} X^{\text{E}} + \bar{g}_{\text{E}}(Z^{\text{E}}) + \epsilon_y^{\text{E}},  \quad (X^{\text{E}}, Z^{\text{E}}) \perp\!\!\!\perp \epsilon_y^{\text{E}}, \label{eq:model_external}
\end{align}
where $\epsilon_y$ and $\epsilon_y^{\text{E}}$ are mean-zero random variables with variances $0 < \sigma^2 < \infty$ and $0 < \sigma_{\text{E}}^2 < \infty$, respectively. The nuisance functions $\bar{g}$ and $\bar{g}_{\text{E}}$ are allowed to differ between cohorts. We consider local alternatives of the form $c = t/\sqrt{n}$ for a fixed $t > 0$. For the external data, we set $c^{\text{E}} = t/\sqrt{f_n}$, where $f_n/n \to \zeta > 0$. We assume the external sample size satisfies $n_{\text{E}}/n \to \tau \in (0, \infty)$.

\begin{theorem}\label{power of unknown cd}
    Consider models \eqref{primary internal data model}--\eqref{unlabel data model}, \eqref{eq:model_internal}, and \eqref{eq:model_external}. Suppose the conditions in Theorem~S.2 in Supplementary Materials~S.4 hold, excluding that under $H_0$. In addition, assume the following conditions hold for the internal data:
    \begin{align}
        &\mathbb{E}(\epsilon_y^4) < \infty, \quad 0 < (Y - \mathbb{E}(Y|Z))^2 < \infty \text{ almost surely}, \label{moment condition 4}\\
        &\frac{1}{|\mathcal{I}_{\mathrm{I}}|}\sum_{i\in \mathcal{I}_{\mathrm{I}}} \left( \widehat{g}(Z_i) - \mathbb{E}(Y|Z_i) \right)^2 = o_{\mathbb{P}}(1), \quad \frac{1}{|\mathcal{I}_{\mathrm{I}}|}\sum_{i\in \mathcal{I}_{\mathrm{I}}} \left( \widehat{\mu}(Z_i) - \mu(Z_i) \right)^2 = o_{\mathbb{P}}(1), \label{con of ghat 1 1} \\
        &\frac{1}{|\mathcal{I}_{\mathrm{I}}|}\sum_{i\in \mathcal{I}_{\mathrm{I}}} \left( \widehat{g}(Z_i) - \mathbb{E}(Y|Z_i) \right)^2 \cdot \frac{1}{|\mathcal{I}_{\mathrm{I}}|}\sum_{i\in \mathcal{I}_{\mathrm{I}}} \left( \widehat{\mu}(Z_i) - \mu(Z_i) \right)^2 = o_{\mathbb{P}}(|\mathcal{I}_{\mathrm{I}}|^{-1}), \label{con of ghat 1 2} \\
        &\frac{1}{|\mathcal{I}_{\mathrm{I}}|}\sum_{i\in \mathcal{I}_{\mathrm{I}}} \left( \widehat{g}(Z_i) - \mathbb{E}(Y|Z_i) \right)^2 \operatorname{Var}_{H_{1n}}(\epsilon_i|Y_i,Z_i) = o_{\mathbb{P}}(1), \label{con of ghat 1 3}\\
        &\frac{1}{\sqrt{|\mathcal{I}_{\mathrm{I}}|}}\sum_{i\in \mathcal{I}_{\mathrm{I}}} \left( \widehat{g}(Z_i) - \mathbb{E}(Y|Z_i) \right) \mathbb{E}_{H_{1n}}(\epsilon_i|Y_i,Z_i) = o_{\mathbb{P}}(1), \label{con of ghat 1 4}
    \end{align}
    and assume the external data satisfy corresponding analogues. If $|\mathcal{I}_{\mathrm{I}}|/n \to w_1$, $|\mathcal{I}_{\mathrm{E}}|/n_{\mathrm{E}} \to w_2$, and $w \to w^*$ under the local alternative, then the local asymptotic power of $\mathrm{CRT}^*$ using $|T_{\mathrm{comb}}|$ is
    \begin{equation*}
        \Phi(\xi - z_{1-\alpha/2}) + \Phi(-\xi - z_{1-\alpha/2}), \quad \text{where} \quad \xi = \sqrt{w_1}t \frac{(1-w^*)\widetilde{\sigma} + w^*\widetilde{\sigma}/\sqrt{\zeta}}{\sqrt{(1-w^*)^2\sigma^2 + \frac{(w^*)^2 w_1 \sigma_{\mathrm{E}}^2}{w_2 \tau}}}.
    \end{equation*}
\end{theorem}

The definition of the local asymptotic power and  clarifications of the conditions are provided in Supplementary Materials~S.5 and~S.6.2, respectively. Theorem~\ref{power of unknown cd} (proof in Supplementary Materials~S.11) reveals several key insights: (i) power increases with $t$ and decreases with $\sigma^2$ and $\sigma_{\text{E}}^2$; (ii) a larger $\tau$ and a smaller $\zeta$ both contribute to increased power; (iii) larger values of $w_1$ and $w_2$ improve power; and (iv) the distilled statistics scale with $\widetilde{\sigma}^2$ (the noise variance of $X|Z$), which leads to higher power as $\widetilde{\sigma}^2$ increases. Further details are provided in Supplementary Materials S.6.3.

Theorem~\ref{power of unknown cd} establishes the first local-power result for CI testing that jointly leverages internal, external, and unlabeled data when conditional distributions are unknown. A key step in the analysis is showing that resampling from the SRB-estimated conditional distributions achieves the same local asymptotic power as resampling from the true conditional distributions. A crucial component of our proof is the additivity property of conditional weak convergence for conditionally independent summands (Lemma S.4 in Supplementary Materials S.11), which generalizes the classical additivity property under independence \citep{van2000asymptotic}. Our analysis extends previous work \citep{wang2022high, katsevich2022power, niu2024reconciling} by accommodating high-dimensional partial linear models without parametric noise assumptions, permitting in-sample nuisance fitting under conditions \eqref{con of ghat 1 1}--\eqref{con of ghat 1 4}, and allowing the covariate dimension $p$ to grow with $n$.

\noindent\textbf{Optimal Convex Combination of Distilled Statistics.} The optimal weight $w^{\text{opt}}$ that maximizes the power in Theorem~\ref{power of unknown cd} is given by:
\begin{equation}\label{optimal weight unknown cd}
    w^{\text{opt}} = \frac{\sigma^2/\sqrt{\zeta}}{w_1\sigma_{\text{E}}^2/(w_2\tau) + \sigma^2/\sqrt{\zeta}}.
\end{equation}
Plugging $w^{\text{opt}}$ into the power expression yields the maximal power:
\begin{equation}\label{maximal power}
    \Phi\left(\sqrt{w_1}t \sqrt{ \frac{\widetilde{\sigma}^2}{\sigma^2} + \frac{\tau w_2\widetilde{\sigma}^2}{\zeta w_1\sigma_{\text{E}}^2} }-z_{1-\alpha/2}\right) + \Phi\left(-\sqrt{w_1}t \sqrt{ \frac{\widetilde{\sigma}^2}{\sigma^2} + \frac{\tau w_2\widetilde{\sigma}^2}{\zeta w_1\sigma_{\text{E}}^2} }-z_{1-\alpha/2}\right). 
\end{equation}
This maximal power always exceeds the power of the test without external data (i.e., where $w = w^* = 0$):
\begin{equation*}
    \Phi\left(\sqrt{w_1}t \frac{\widetilde{\sigma}}{\sigma}-z_{1-\alpha/2}\right) + \Phi\left(-\sqrt{w_1}t \frac{\widetilde{\sigma}}{\sigma}-z_{1-\alpha/2}\right).
\end{equation*}
The power gain is monotonically increasing with the external sample size (larger $\tau$) and the strength of the external conditional dependence (smaller $\zeta$). For a detailed comparison between $w^{\text{opt}}$ and the optimal weights typically used for parameter estimation in the data fusion literature \citep{lin2010relative, hu2022paradoxes}, see Supplementary Materials~S.7.

We now show how to consistently estimate this optimal weight. Theorem~\ref{power of CRT*OPDS unknown cd} establishes that a weight based on the estimated residual variances is consistent under the local alternative.

\begin{theorem}\label{power of CRT*OPDS unknown cd}
    Under models \eqref{primary internal data model}--\eqref{unlabel data model}, \eqref{eq:model_internal}, and \eqref{eq:model_external}, suppose $\max\left\{\mathbb{E}(\epsilon_y^4), \mathbb{E}[(\epsilon_y^{\mathrm{E}})^4]\right\} < \infty$ and the nuisance estimators satisfy the consistency condition:
    \begin{small}
    \begin{equation}\label{con of ghat and gEhat 2}
        \max\left\{ \frac{1}{|\mathcal{I}_{\mathrm{I}}|}\sum_{i\in \mathcal{I}_{\mathrm{I}}} (\widehat{g}(Z_i) - \mathbb{E}(Y|Z_i))^2,\ \frac{1}{|\mathcal{I}_{\mathrm{E}}|}\sum_{i\in \mathcal{I}_{\mathrm{E}}} (\widehat{g}_{\mathrm{E}}(Z^{\mathrm{E}}_i) - \mathbb{E}(Y^{\mathrm{E}}|Z^{\mathrm{E}}_i))^2 \right\} = o_{\mathbb{P}}(1).
    \end{equation}
    \end{small}%
    Then, the \textbf{inverse-residual weight}:
    \begin{equation*}
        \widehat{w} = \frac{S_{\mathrm{int}}/\sqrt{\zeta}}{ S_{\mathrm{ext}} / (|\mathcal{I}_{\mathrm{E}}|/|\mathcal{I}_{\mathrm{I}}|) + S_{\mathrm{int}}/\sqrt{\zeta} },
    \end{equation*}
    where $S_{\mathrm{int}} = |\mathcal{I}_{\mathrm{I}}|^{-1}\sum_{i\in \mathcal{I}_{\mathrm{I}}} (Y_i - \widehat{g}(Z_i))^2$ and $S_{\mathrm{ext}} = |\mathcal{I}_{\mathrm{E}}|^{-1}\sum_{i\in \mathcal{I}_{\mathrm{E}}}(Y^{\mathrm{E}}_i - \widehat{g}_{\mathrm{E}}(Z^{\mathrm{E}}_i))^2$, is a \textbf{consistent estimator} of $w^{\mathrm{opt}}$ under the local alternative. The resulting combined test statistic is 
    \begin{equation}\label{IRW stat}
        \widehat{T}_{\mathrm{comb}} = (1-\widehat{w}) T_{\mathrm{int}} + \widehat{w} T_{\mathrm{ext}}.
    \end{equation} Under the conditions of Theorem~\ref{power of unknown cd}, the local asymptotic power of $\mathrm{CRT^*}$ based on $|\widehat{T}_{\mathrm{comb}}|$ achieves the maximum power defined in \eqref{maximal power}.
\end{theorem}

The proof is provided in Supplementary Materials~S.11. With this data-driven weighting scheme, CRT* adaptively incorporates external data even under cohort heterogeneity, providing a principled way to boost power in high-dimensional regimes where internal samples are limited. 

We further address extensions of this framework in the Supplementary Materials, including cases where noise terms are not identically distributed (Section S.8) and a novel data-driven procedure to identify the informative sets $\mathcal{A}$ and $\mathcal{A}^{\text{E}}$ when they are unknown (Section S.9).

\section{Simulation Studies}\label{simulation results}

\subsection{Evaluated Methods and Test Statistics}\label{sec:Evaluated Methods}

We compare two CRT* training strategies: (i) \textbf{Hold-out training}, which splits labeled data for conditional distribution estimation and test statistic computation;
and (ii) \textbf{In-sample training}, which uses all labeled data for both steps.
The full pseudocode is in Supplementary Materials~S.12 (Algorithms S.2 and S.3). We use $|\widehat{T}_{\text{comb}}|$ as the test statistic. Here, $\widehat{g}$ and $\widehat{\mu}$, as well as $\widehat{g}_{\text{E}}$ and $\widehat{\mu}_{\text{E}}$, are estimated using the Lasso method.
We refer to this statistic as ``ours''. For comparison, ``pool" pools the internal and external data together and computes the test statistic using the pooled data. Detailed definitions of ``ours" and ``pool" are provided in Supplementary Materials~S.13.1. We  focus on the case with known informative sets. Results for the unknown informative set are in Supplementary Materials~S.13.7. All experiments use a significance level $\alpha=0.05$.

\subsection{Simulation Setup}\label{sec: sim setup}

To evaluate the performance of CRT*, we generate internal, external, and auxiliary unlabeled datasets with a covariate dimension of $p=200$. The covariates $Z$, $Z^{\text{E}}$, and $Z^{(k)}$ ($k=1,\ldots,K$ with $K=12$) are drawn i.i.d. from $\mathcal{N}(\mathbf{0}, \mathbf{I}_{p})$. The predictors $X$, $X^\text{E}$, and $X^{(k)}$  are  generated according to models \eqref{primary internal data model}, \eqref{primary external data model}, and \eqref{unlabel data model}. The response $Y$ is generated via \eqref{eq:model_internal}, where we set $c = 0$ to evaluate type-I error control and vary $c > 0$ for power assessment. The external response $Y^{\text{E}}$ is generated similarly via \eqref{eq:model_external}.

We consider three data-generating scenarios: (i) \textbf{DS}: different models across cohorts but the same conditional dependence (CD) strength; (ii) \textbf{DD}: different models and different CD strengths; and (iii) \textbf{SS}: same model and CD strength. We focus on results for scenarios DS and DD here. The detailed data-generating mechanisms and results for scenario SS are provided in Supplementary Materials~S.13.2.1 and S.13.2.3, respectively.

For \textit{in-sample training}, we set the internal sample size to $n = 100$ and vary the external sample size $n_{\text{E}} \in \{0, 100, 200, 300\}$. To evaluate the impact of unlabeled data on type-I error, we vary the size of each unlabeled dataset $n_0 \in \{0, 50, 100, 150, 200\}$ (yielding $N = 12n_0$ total unlabeled samples). For  power analysis, we fix $n_0 = 200$ to ensure accurate conditional distribution estimation. For \textit{hold-out training}, the internal and external sample sizes are doubled, and each dataset is split equally between distribution estimation and test statistic computation. 

The special cases where $n_{\text{E}}=0$ (no external data) and/or $n_0=0$ (no unlabeled data) are handled as described in Supplementary Materials~S.13.3. All CRT* $p$-values are computed using $M=200$ resamples. Type-I error and power are calculated as averages over 1000 independent replications. Detailed tuning parameter selection is described in Supplementary Materials~S.13.4.

\begin{figure}[htbp]
\centering
\begin{minipage}{0.96\linewidth}
    \centering
    {\small (a) Hold-out training} \\
    \includegraphics[width=\textwidth]{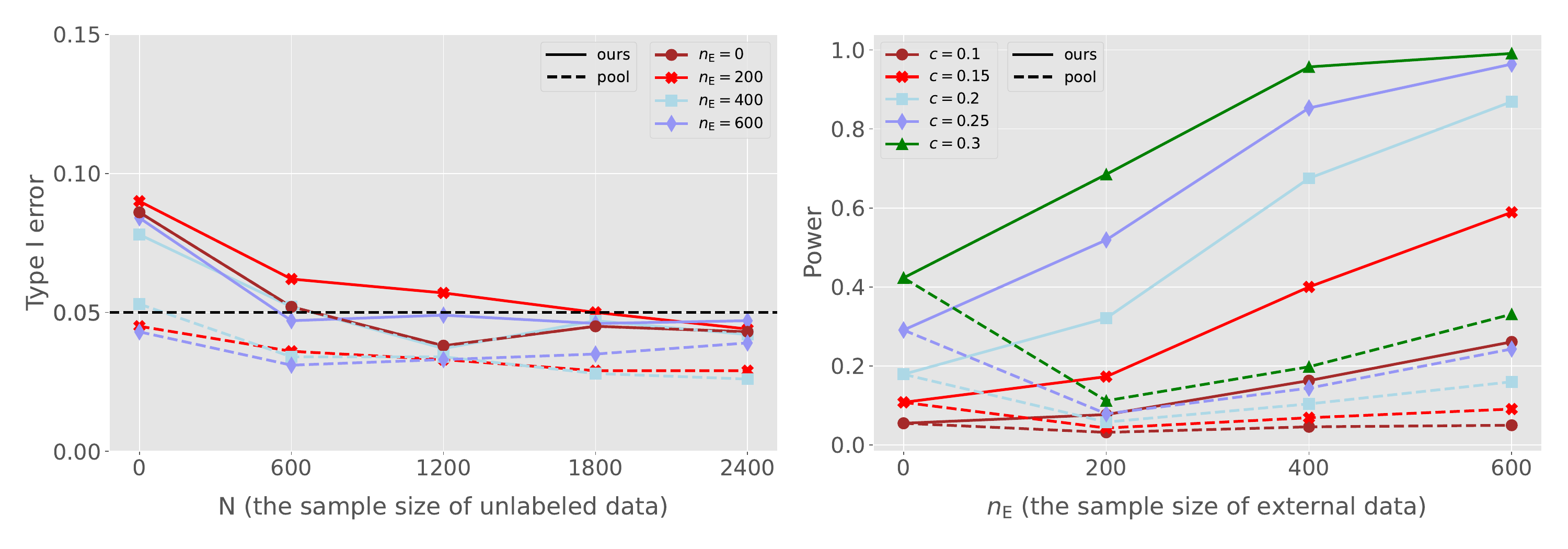}
\end{minipage}
\begin{minipage}{0.96\linewidth}
    \centering
    {\small (b) In-sample training} \\
    \includegraphics[width=\textwidth]{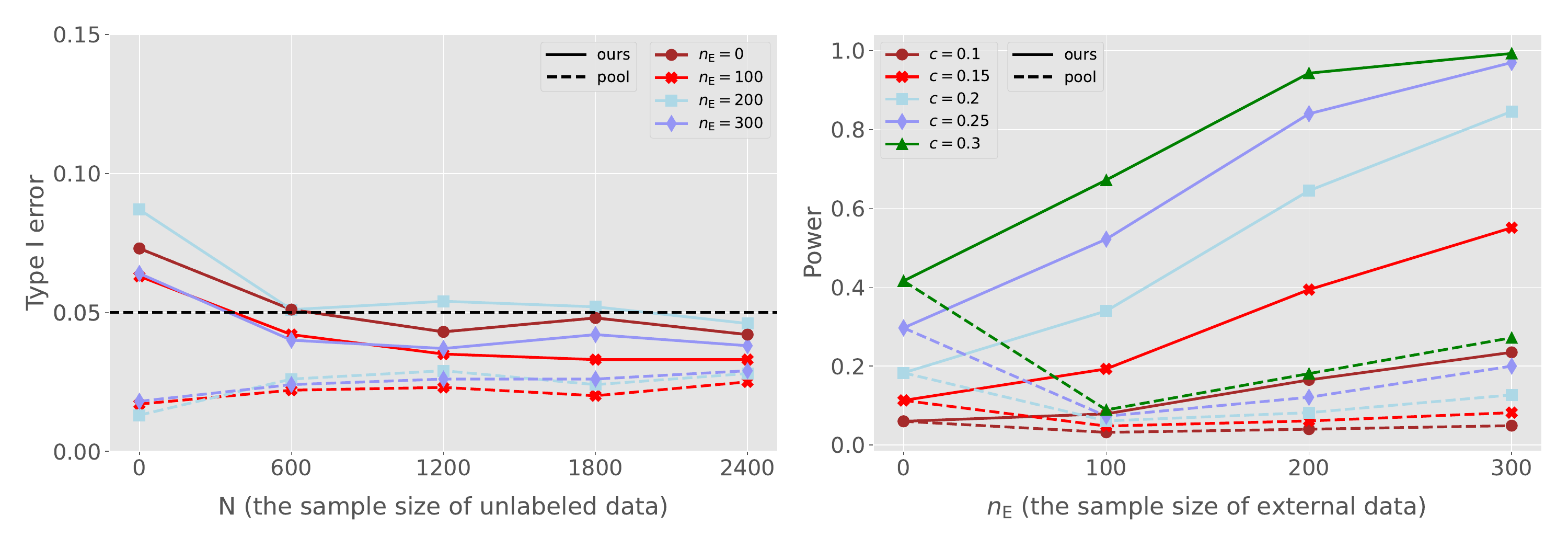}
\end{minipage}
\caption{Type-I error (left) and power (right) for CRT* under hold-out and in-sample training in setting (DS.I). In the type-I error plots, the x-axis $N=12n_0$ denotes the total unlabeled sample size. Colors indicate different external sample sizes ($n_{\text{E}}$) for type-I error and different conditional dependence strengths ($c$) for power. Solid lines represent our proposed adaptive statistic (``ours''), while dashed lines represent the pooled statistic (``pool''). In the type-I error plots, when $n_{\text{E}}=0$, the solid and dashed lines overlap, because  ``ours” and ``pool” statistics are identical. 
The nominal level $\alpha=0.05$ is indicated by the horizontal dashed line.} 
\label{fig:DS.I}
\end{figure}

\subsection{Different Models but Same CD Strength (DS)} \label{KIS,DM,SC}

We first consider settings where the internal and external datasets are generated from different models (exhibiting distributional heterogeneity) but share the same conditional dependence (CD) strength. 
We evaluate two settings, (DS.I) and (DS.II), with specifications detailed in Supplementary Materials~S.13.2.1.

The results for (DS.I) are shown in Figure~\ref{fig:DS.I}, while those for (DS.II) are provided in Supplementary Materials~S.13.2.2 (Figure~S.1). Key findings, which are consistent across both hold-out and in-sample training strategies, include:

{\setlength{\leftmargini}{1em}
\begin{itemize}
    \item \textbf{Type-I Error Control:} Type-I error remains well-controlled at the nominal level for both the ``pool'' and ``ours'' statistics, provided the unlabeled sample size is sufficiently large to ensure accurate conditional distribution estimation.
    \item \textbf{Failure of Naive Pooling:} Under the alternative ($H_1$), the power of the ``pool'' statistic does not increase monotonically with the external sample size $n_{\text{E}}$. In fact, the power initially deteriorates when $n_{\text{E}}$ is small. This occurs because naive pooling fails to adapt to distributional heterogeneity; the resulting distilled statistic is biased toward zero, leading to a significant loss of power. While the power recovers slightly as $n_{\text{E}}$ becomes large enough for the external data to dominate the pooled sample, it remains substantially lower than that of our proposed method.
    \item \textbf{Adaptive Fusion Gains:} Unlike naive pooling, our method effectively leverages external data to improve power even with a small $n_{\text{E}}$. Consistent with our theoretical results, the power gain is more pronounced as the external sample size increases. 
    By separately estimating statistics and combining them optimally, CRT* accommodates cohort-specific distributions and bypasses the biases inherent in naive pooling.
\end{itemize}}

\subsection{Different Models and Different CD Strengths (DD)} \label{KIS,DM,DC}

We next examine settings where the cohorts are generated from distinct models with different CD strengths, specifically (DD.I) and (DD.II), as detailed in Supplementary Materials~S.13.2.1. For type-I error evaluation, the conditional mutual information (CMI) in the external data is set to be small, as required by Theorem~\ref{asymptotic valid SSL-CRT1}. For power evaluation, we vary the strength of the conditional dependence.

The results for settings (DD.I) and (DD.II) are presented in Figures~S.2 and S.3, respectively (Supplementary Materials~S.13.2.2). The findings across both training strategies are summarized below:

{\setlength{\leftmargini}{1em}
\begin{itemize}
    \item \textbf{Robustness of Type-I Error:} Type-I error is effectively controlled when the unlabeled sample size is large. Crucially, control is maintained even when the external data only approximately satisfy conditional independence under the null, consistent with our theory.
    \item \textbf{Failure of Naive Pooling:} Similar to the DS scenario, the power of ``pool'' typically suffers under small $n_{\text{E}}$ and fails to match the performance of our adaptive approach.
    \item \textbf{Power Enhancement:} Our method consistently improves power by incorporating external data, even when the external cohort exhibits weaker conditional dependence than the internal cohort. The power
gain is more pronounced when the external sample size is larger and/or when the conditional dependence in the external data
is stronger.
\end{itemize}}

\subsection{Additional Important Findings}

The detailed experimental results in Supplementary Materials~S.13.5--S.13.7 yield several critical insights into the robustness and necessity of the CRT* framework:

{\setlength{\leftmargini}{1em}
\begin{itemize}
    \item \textbf{Robustness to Non-Gaussianity:} SRB-based CRT* maintains valid type-I error control even with non-Gaussian residuals, whereas Gaussian approximation sampling (GAS) fails (see Supplementary Materials~S.13.5). This robustness stems from the fact that SRB approximates the noise distribution nonparametrically, making it far less sensitive to residual-distribution misspecification than parametric alternatives.
    
    \item \textbf{Necessity of Smoothing:} The unsmoothed residual-bootstrap fails to control type-I error (see Supplementary Materials~S.13.6). This confirms our theoretical assertion: without smoothing, the bootstrap distribution remains discrete. This prevents the estimated conditional distribution from converging in TV distance to the true (continuous) distribution, leading to invalid $p$-values.
    
    \item \textbf{Trans-Lasso vs. Naive Pooling:} When the informative set is unknown, CRT* using SRB with Trans-Lasso maintains valid type-I error control when the unlabeled sample size is large and consistently enhances power. Conversely, naively pooling internal data with unlabeled data to estimate $X|Z$, and external data with unlabeled data to estimate $X^{\mathrm{E}}|Z^{\mathrm{E}}$, can induce mismatches between the estimated and true conditional distributions, leading to inflated type-I error (see Supplementary Materials~S.13.7). 
\end{itemize}}

In conclusion, our proposed CRT* provides a principled framework for CI testing under distributional heterogeneity. 
CRT* ensures valid type-I error control, even when external data exhibit weak conditional dependence under the null. Under the alternative, the procedure achieves consistent power gains by adaptively incorporating external data, even when cohort models or CD strengths differ. In contrast, naive pooling strategies often fail to either control the type-I error or provide any meaningful improvement in testing power.

\begin{table}[htbp]
\centering
\small
\setlength{\tabcolsep}{8pt}
\renewcommand{\arraystretch}{1.1}
\begin{tabular}{lllc}
\toprule
\textbf{Data Type} & \textbf{Platform} & \textbf{Population/Subtype} & \textbf{Sample Size} \\
\midrule
Internal Data  & UCSC Xena & African American & 187 \\
External Data  & UCSC Xena & White           & 680 \\
Unlabeled Data & GEO       & Basal           & 360 \\
               &           & Her2            & 348 \\
               &           & LumA            & 1709 \\
               &           & LumB            & 767 \\
               &           & Normal-like     & 225 \\
\bottomrule
\end{tabular}
\caption{Summary of BRCA datasets.}
\label{tab:BRCA data}
\end{table}
\section{Real Data Analysis}\label{real data}

\subsection{Data and Scientific Problem}
We demonstrate the effectiveness of CRT* using RNA-seq gene expression data from breast cancer (BRCA) patients, the most common cancer diagnosis among women worldwide. We obtained data from the TCGA-BRCA cohort via the UCSC Xena platform, comprising 1,227 samples across 60,660 genes, and from the Gene Expression Omnibus (GEO) database, comprising 3,409 samples across 30,865 genes.

Our objective is to investigate the potential regulatory relationship between \textit{BRCA1}, a well-known tumor suppressor gene, and its pseudogene \textit{BRCA1P1}, which has been reported to express a long non-coding RNA associated with breast cancer \citep{han2021brca1}. Specifically, we aim to determine if an edge exists between \textit{BRCA1P1} ($Y$) and \textit{BRCA1} ($X$) in the undirected gene regulatory network for the African American TCGA cohort. This requires testing whether $X$ and $Y$ are conditionally independent given the expression of other genes ($Z$).

For this analysis, we treat African American TCGA patients as the internal cohort ($n=187$) and White TCGA patients as the external cohort ($n_{\text{E}}=680$). The GEO samples lack expression data for the pseudogene \textit{BRCA1P1} but provide a significantly larger sample size for the other genes, and are thus treated as unlabeled data. The GEO samples are stratified into five Prediction Analysis of Microarray 50 (PAM50) subtypes: Basal, Her2, LumA, LumB, and Normal-like. After preprocessing (detailed in Supplementary Materials~S.14.1), we select the 200 most differentially expressed genes. We retain \textit{BRCA1} as the predictor $X$ and treat the remaining 199 genes as the conditioning variables $Z$. A summary of the datasets is provided in Table~\ref{tab:BRCA data}. Diagnostic analyses (Supplementary Materials~S.14.2) confirm significant distributional differences across the internal, external, and unlabeled cohorts. All subsequent tests are performed at a significance level of $\alpha = 0.05$.

\subsection{Performance of CRT*}

Since the informative set is unknown in this application, we implement CRT* using Trans-Lasso (detailed in Supplementary Materials~S.9) combined with the SRB procedure. The resulting algorithms are summarized in Algorithms~S.7 and S.8 (Supplementary Materials~S.12).

We first perform CI testing separately for the internal and external cohorts using in-sample training. Each analysis incorporates the auxiliary unlabeled data (i.e., internal + unlabeled, or external + unlabeled) and uses the absolute value of the distilled Lasso statistic as the test statistic (see Supplementary Materials~S.14.3 for details). The resulting $p$-values are $0.015$ for the internal cohort and $0.045$ for the external cohort, providing preliminary evidence against CI in both populations. Furthermore, one-sided tests yield $p$-values of $0.015$ (internal) and $0.040$ (external), with positive statistics in both cases, suggesting a positive conditional association between \textit{BRCA1} and \textit{BRCA1P1}.

By incorporating the external data using our proposed in-sample CRT* (Algorithm~S.8) with the adaptive ``ours'' statistic (see Simulation Studies), we obtain a $p$-value of \textbf{0.010}. This is smaller than the $p$-value obtained using only the internal and unlabeled data, demonstrating that CRT* successfully leverages external information to provide stronger evidence against the null hypothesis.

\subsection{Comparisons with Naive and Existing Methods under In-Sample Training}

We compare our proposed in-sample CRT* against two naive CRT* variants (``Pool Stat'' and ``Pool SRB'') and three state-of-the-art methods: CRT, CPT \citep{berrett2020conditional}, and  in-sample Maxway$_{\text{in}}$ CRT \citep{li2023maxway}. The variants are defined as follows (detailed implementations are  in Supplementary Materials S.14.3):

{\setlength{\leftmargini}{1em}
\begin{itemize}
    \item \textbf{``Pool Stat'':} Replaces our adaptive fusion statistic (``ours") with the pooled statistic (``pool") used in Simulation Studies.
    \item \textbf{``Pool SRB'':} Internal data are pooled with unlabeled data to estimate $X|Z$, while external data are pooled with unlabeled data to estimate $X^{\mathrm{E}}|Z^{\mathrm{E}}$, 
    ignoring potential heterogeneity.
    \item \textbf{``Internal'' vs. ``Pool'':} For existing methods, ``Internal'' uses only the internal cohort as the primary dataset, while ``Pool'' naively combines the internal and external cohorts into the primary dataset.
\end{itemize}}
The results are summarized in Table~\ref{tab:pvalues-for-real_data}. Our method achieves the most significant $p$-value (\textbf{0.010}). Notably, the ``Pool Stat'' variant fails to reject the null at the $\alpha=0.05$ level ($p=0.055$), highlighting that naive pooling can diminish power under heterogeneity. Existing methods using only the internal and unlabeled data (``Internal'') show higher $p$-values due to limited sample sizes, while their ``Pool'' implementations show even weaker evidence against the null, consistent with the power loss phenomenon discussed in Section~\ref{subsec:crt*}.


\begin{table}[ht]
\centering
\small
\renewcommand{\arraystretch}{1.2}
\begin{adjustbox}{max width=\textwidth}
\begin{tabular}{lccccccccc}
\toprule
\multirow{2}{*}{Method} & \multirow{2}{*}{\textbf{Ours}} & \multirow{2}{*}{Pool Stat} & \multirow{2}{*}{Pool SRB} & \multicolumn{2}{c}{CRT} & \multicolumn{2}{c}{CPT} & \multicolumn{2}{c}{Maxway$_{\text{in}}$ CRT} \\
\cmidrule(lr){5-6} \cmidrule(lr){7-8} \cmidrule(lr){9-10}
& & & & Internal & Pool & Internal & Pool & Internal & Pool \\
\midrule
$p$-value & \textbf{0.010} & 0.055 & 0.015 & 0.020 & 0.045 & 0.030 & 0.035 & 0.011 & 0.023 \\
\bottomrule
\end{tabular}
\end{adjustbox}
\caption{$p$-values for testing CI between \textit{BRCA1} and \textit{BRCA1P1} across different methods and data-pooling strategies.}
\label{tab:pvalues-for-real_data}
\end{table}
\begin{table}[ht]
\centering
\small
\renewcommand{\arraystretch}{1.2}
\begin{adjustbox}{max width=\textwidth}
\begin{tabular}{lcccccc}
\toprule
\multirow{2}{*}{Method} & \multirow{2}{*}{\textbf{Ours}} & \multirow{2}{*}{Internal} & \multirow{2}{*}{Pool Stat} & \multirow{2}{*}{Pool SRB} & \multicolumn{2}{c}{Maxway$_{\text{out}}$ CRT} \\
\cmidrule(lr){6-7}
& & & & & Internal & Pool \\
\midrule
Rejection rate & \textbf{87\%} & 43\% & 33\% & 80\% & 48\%  & 19\% \\
\bottomrule
\end{tabular}
\end{adjustbox}
\caption{Rejection rates across 100 random splits for CI testing between \textit{BRCA1} and \textit{BRCA1P1} under hold-out training.}
\label{tab:rt-for-real_data}
\end{table}
\subsection{Comparisons with Naive and Existing Methods under Hold-Out Training}

We further compare our proposed hold-out training procedure, \textbf{``Ours''} (Algorithm~S.7 with the adaptive fusion statistic), against three naive hold-out CRT* variants: (i) \textbf{``Internal''}, which utilizes only the internal and unlabeled data (see Supplementary Materials~S.14.3); (ii) \textbf{``Pool Stat''}; and (iii) \textbf{``Pool SRB''}. We also include the hold-out version of Maxway CRT (Maxway$_{\text{out}}$ CRT) in both its ``Internal'' and ``Pool'' implementations. Since the original CRT and CPT methods do not natively support hold-out training for conditional distribution estimation, they are excluded from this specific comparison. 

Because hold-out procedures require sample splitting, we randomly partition both the internal and external samples into two equal halves (one for distribution estimation and one for test statistic computation) and repeat this process 100 times. Table~\ref{tab:rt-for-real_data} reports the rejection rates at the $\alpha=0.05$ level across these splits.
Our analysis yields several key observations:
{\setlength{\leftmargini}{1em}
\begin{itemize}
    \item \textbf{Power Enhancement:} ``Ours'' achieves the highest rejection rate (\textbf{87\%}). In contrast, the ``Internal'' versions of CRT* and Maxway$_{\text{out}}$ CRT reject in only 43\% and 48\% of splits, respectively. This discrepancy highlights the significant power loss induced by sample splitting when the internal dataset is small, a limitation that our method overcomes by effectively incorporating external data. Notably, in 74\% of all splits, ``Ours'' produces a smaller $p$-value than the ``Internal'' CRT*, and in 91\% of the cases where the ``Internal'' CRT* failed to reject the null, our method successfully identified the conditional association.
    \item \textbf{Failure of Naive Pooling:} When Maxway$_{\text{out}}$ CRT is implemented using naive pooling, its rejection rate drops precipitously to 19\%. This indicates that instead of gaining power, the test becomes less powerful than its ``Internal" counterpart due to unaddressed cohort heterogeneity.
    \item \textbf{Importance of Heterogeneity Modeling:} Both the ``Pool Stat'' and ``Pool SRB'' variants of CRT* reject less frequently than our proposed adaptive method, confirming that accounting for cross-cohort differences is essential when leveraging multi-source data.
\end{itemize}}



\subsection{Summary of Biological Findings}

This comprehensive analysis demonstrates the  power gains achieved by the proposed framework. By adjusting for the expression of other genes in a high-dimensional regime, our study provides robust statistical evidence of a regulatory edge between \textit{BRCA1P1} and \textit{BRCA1} in the African American breast cancer cohort. This finding is supported by and further complements existing biomedical literature, in which \textit{BRCA1P1} has been reported to be involved in recombination or gene-conversion events at the \textit{BRCA1} locus \citep{puget2002distinct,tessereau2015occurrence}.
The ability of CRT* to identify this relationship---where methods using only the internal and unlabeled data or naively pooling heterogeneous data may fail---underscores its practical utility for feature selection and causal inference in precision medicine.

\section{Conclusion}
\label{conclusion}
We propose a novel CRT-based framework for high-dimensional CI testing that effectively integrates external and unlabeled data. In the presence of distributional heterogeneity, our method theoretically guarantees valid type-I error control under the null, even when the external dataset exhibits weak conditional dependence. Moreover, it achieves consistent power improvements under the alternative, even when the strengths of conditional dependence differ between the internal and external datasets. Extensive simulations and real data analyses demonstrate the robustness and superior performance of our approach. Therefore, our proposed framework has the potential to leverage multi-source datasets to enhance causal discovery in real-world applications, such as gene regulatory networks or complex social networks, and to more effectively uncover relationships and patterns in complex systems.

\bibliographystyle{abbrvnat}
\bibliography{reference}

\end{document}